\documentclass[reprint,twocolumn,showpacs,superscriptaddress,longbibliography,email,floatfix,aps,prl]{revtex4-2}

\usepackage{header}

\begin{document}
\title{Thermodynamics of quantum trajectories on a quantum computer}
\author{Marcel Cech}
\affiliation{Institut f\"ur Theoretische Physik, Universit\"at T\"ubingen, Auf der Morgenstelle 14, 72076 T\"ubingen, Germany}
\author{Igor Lesanovsky}
\affiliation{Institut f\"ur Theoretische Physik, Universit\"at T\"ubingen, Auf der Morgenstelle 14, 72076 T\"ubingen, Germany}
\affiliation{School of Physics and Astronomy and Centre for the Mathematics and Theoretical Physics of Quantum Non-Equilibrium Systems, The University of Nottingham, Nottingham, NG7 2RD, United Kingdom}
\author{Federico Carollo}
\affiliation{Institut f\"ur Theoretische Physik, Universit\"at T\"ubingen, Auf der Morgenstelle 14, 72076 T\"ubingen, Germany}

\begin{abstract}
    Quantum computers have recently become available as noisy intermediate-scale quantum devices. Already these machines yield a useful environment for research on quantum systems and dynamics. Building on this opportunity, we investigate open-system dynamics that are simulated on a quantum computer by coupling a system of interest to an ancilla. After each interaction the ancilla is measured and the sequence of measurements defines a quantum trajectory. Using a thermodynamic analogy, which identifies trajectories as microstates, we show how to  bias  the dynamics of the open system in order to enhance the probability of quantum trajectories with desired properties, e.g., particular  measurement patterns or temporal correlations. We discuss how such  a biased --- generally non-Markovian --- dynamics can be implemented on a unitary, gate-based quantum computer and show proof-of-principle results on the publicly accessible \texttt{ibmq\_jakarta} machine. While our study is solely conducted on small systems, it highlights the challenges in controlling complex aspects of open-system dynamics on digital quantum computers.
\end{abstract}

\maketitle


\textit{Introduction.--- } A widely pursued quest in contemporary physics research concerns the realization of a universal quantum computer. However, while fundamental issues --- such as scalability and the necessity of error correction --- have been identified \cite{DiVincenzo2000,Preskill2021,BlumeKohout2002,Nielsen2010,Preskill2018}  the availability of a fault-tolerant machine, which is able to implement arbitrary unitary circuits, may still be years away. Currently available quantum computation platforms belong to the class of noisy intermediate-scale devices \cite{Preskill2018,Lau2022,Leymann2020}. Nevertheless, these machines constitute intriguing systems for conducting proof-of-principle studies, for testing the efficient implementation of quantum algorithms and for making conceptual progress in the understanding of the utility of quantum computers \cite{Feynman1982,Arute2019,Keenan2022,Friedman2022,Daley2022,Kim2021,Corcoles2021,Buluta2009,Francis2021,Leontica2021,Xiao2021,Smith2019,GarciaPerez2020,Chisholm2022,Cattaneo2023}.

\begin{figure}[t]
    \centering
    \includegraphics{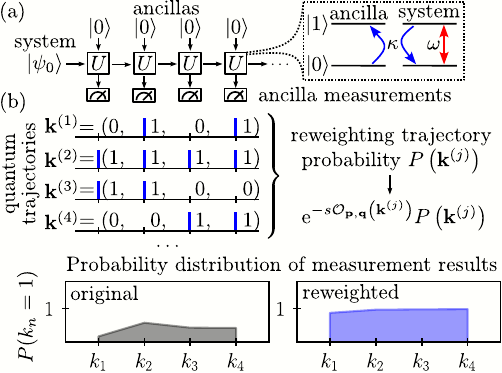}
    \caption{\textbf{Open-system dynamics and quantum trajectories in the collision model.} 
    (a) The two-level system, in the initial state $\ket{\psi_0}$, collides sequentially with ancillas (also two-level systems) all prepared in the $\ket{0}$ state. During a collision the system and an ancilla interact via an exchange interaction visualized in the inset, see also Eq.\,(\ref{eq:OriginalCollisionHamiltonian}).
    (b) The outcomes $k_n$ of projective measurements on the ancillas produce a quantum trajectory, whose $j$th realization is denoted by $\mathbf{k}^{(j)} = (k_1^{(j)}, \dots, k_N^{(j)})$. This trajectory occurs with a probability $P(\mathbf{k}^{(j)})$. 
    We formally  bias the statistics of the ensemble of trajectories by reweighting the probabilities with respect to the ``energy"-function $\mathcal{O}_{\mathbf{p}, \mathbf{q}}(\mathbf{k})$, where $\mathbf{p}$ and $\mathbf{q}$ are coupling constants and $s$ can be interpreted as ``inverse temperature".
    The probability distribution for observing the outcome $k_n = 1$, when measuring the state of the $n$th ancilla, is shown at the bottom, for the original as well as the reweighted, i.e. biased, ensemble.}
    \label{fig:fig1}
\end{figure}

In this work we focus on open-system dynamics implemented on a quantum computer via unitary gates \cite{GarciaPerez2020,Barreiro2011,Schindler2013,Lu2017,Xin2017,Hu2020,Kamakari2022,Mi2023}. The basis for this approach is the so-called collision model \cite{GarciaPerez2020,Ciccarello2022,Cilluffo2021,Horssen2015,Ziman2005}. Here irreversible open-system dynamics is generated by creating entanglement between the system of interest and a series of ancillary two-level systems, as depicted in Fig.~\ref{fig:fig1}(a). Measuring the ancillas generates so-called quantum trajectories [cf.~Fig.~\ref{fig:fig1}(b)] which carry information on the dynamical evolution of the system. We show how to  bias  the properties of these trajectories, such as the rate of certain measurement outcomes  on the ancillas and their temporal correlations. Our biasing protocol relies on interpreting a trajectory as a microstate of a fictitious spin system. Defining an ``energy"-function similar to that of the Ising model, we  derive  a  dynamics, which enhances or reduces the probability of certain trajectories in close resemblance to the Boltzmann weight of equilibrium statistical mechanics, as highlighted in Fig.~\ref{fig:fig1}(b). We show how such a deformed --- at first glance unphysical ---  probability can be obtained on a quantum computer. Our study broadens the spectrum of use cases for quantum computers within the domain of non-equilibrium quantum systems, and provides first proof-of-principle results on the \texttt{ibmq\_jakarta} machine. 



\textit{Collision model.--- }
We consider a quantum system subject to a Markovian discrete-time collision-model dynamics,  illustrated in Fig.~\ref{fig:fig1}(a). 
The system state is encoded in the pure state $\ket{\psi}$ and, within each discrete time step, it collides with an ancillary two-level system described through the computational basis $\left\{ \ket{0}, \ket{1} \right\}$. We assume that each ancilla is initialized in the reference state $\ket{0}$ and that the system-ancilla interaction is described by a unitary $U = \mathrm{e}^{-ih}$, with $h$ a hermitian operator. After the collision, the ancilla is measured in the computational basis. This results in a stochastic evolution for the system  described in terms of Kraus operators $K_k = \bra{k} U \ket{0}$ \cite{Cilluffo2021,Ciccarello2022}. Indeed, with probability $P_k = \left\| K_k \ket{\psi} \right\|^2$, the measurement outcome $k$ is observed for the ancilla and the system evolves as $\frac{K_k \ket{\psi}}{\sqrt{P_k}}$.
Performing $N$ collisions and measurements yields a quantum trajectory $\mathbf{k} = (k_1, \dots, k_N)$ as shown in Fig.~\ref{fig:fig1}(b). The probability for the occurrence of a specific trajectory is $P(\mathbf{k}) = \left\| K_{k_N} \dots K_{k_1} \ket{\psi_0} \right\|^2$. 
The discrete-time dynamics of the state of the system, averaged over all possible measurement outcomes, is described by the Kraus map $\mathcal{E}[\rho] = \sum_k K_k \rho K_k^\dagger$.

We study here the simplest case in which the system is a two-level system, just like the ancillas. Furthermore, we set 
\begin{align}
    h = \omega \left(\mathds{1} \otimes \sigma_x\right) + \kappa \left( \sigma_+ \otimes \sigma_- + \sigma_- \otimes \sigma_+ \right)\, ,
    \label{eq:OriginalCollisionHamiltonian}
\end{align}
with $\sigma_- = \sigma_+^\dagger = \ketbra{0}{1}$ and $\sigma_x=\sigma_- +\sigma_+$. 
Throughout, we set the parameters to $\omega = \kappa = 1$ and select $\ket{\psi_0} = \ket{0}$ as the initial state of the system. Contrarily to this genuine discrete-time dynamics, by introducing a time unit $\Delta t$ together with the rescaling $\omega \to \omega \Delta t$ and $\kappa \to \sqrt[]{{\kappa}{\Delta t}}$, we note that the system dynamics converges in the limit $\Delta t \to 0$ towards the continuous-time Lindblad equation of a two-level atom with Rabi frequency $\omega$ and decay rate $\kappa$ (see Ref.~\cite{SM} \vphantom{\cite{Garrahan2010,Carollo2018,QiskitNoiseModel2023}} for details). 



\textit{Biased quantum trajectories.--- } 
Given that the ancillas are two-level systems, each quantum trajectory $\mathbf{k} = (k_1, \dots, k_N)$ is a sequence of zeros and ones.  Within a thermodynamic analogy (see also, e.g.,  Refs.~\cite{lecomte2007,touchette2009,Garrahan2009,Chetrite2013,Chetrite2015}),  each of these sequences can be interpreted as a microscopic configuration (microstate) of a fictitious one-dimensional Ising spin system.  The probability over these microstates is then given by $P(\mathbf{k})$. In the standard thermodynamic approach,  $P(\mathbf{k})$ is assumed to be a flat probability \cite{touchette2009}. Here, however, the collision model provides, in general, a non-flat probability $P(\mathbf{k})$, which, however, does not spoil the thermodynamic analogy. Our goal is to reweigh (or bias) this probability by using an ``energy"-function [see Fig.~\ref{fig:fig1}(b)]
\begin{align}
    \mathcal{O}_{\mathbf{p}, \mathbf{q}}(\mathbf{k}) = \sum_{n=1}^N p_n  k_n + \sum_{\underset{n>m}{n, m = 1}}^N q_{nm} k_n k_m\, ,
    \label{eq:GeneralEnergyFunction}
\end{align} 
defined in terms of the set of real coefficients $\left\{p_n, q_{nm}\right\}$.  In analogy with the Ising model, the vector $\mathbf{p}=(p_1, \dots, p_N)$ encodes the interaction of the spin system with a (possibly inhomogeneous) external field, while the matrix $\mathbf{q}$, collecting all the terms $q_{nm}$, describes the two-spin interacting energy. The reweighting (or biasing) is formally achieved by introducing the {\it canonical} Gibbs probability $P(\mathbf{k},s)\propto \mathrm{e}^{-s\mathcal{O}_{\mathbf{p}, \mathbf{q}}(\mathbf{k})} P(\mathbf{k})$, in which $\mathrm{e}^{-s\mathcal{O}_{\mathbf{p}, \mathbf{q}}(\mathbf{k})}$ represents a Boltzmann weight.
Within this thermodynamic construction, the parameter $s$ in the biasing factor serves as an ``inverse temperature” \cite{Garrahan2009,Garrahan2010,Carollo2019}. The probability of the different quantum trajectories is then modified on-demand by tuning the ``temperature", as well as the ``energy"-function, such that the outcome of the measurements tends to minimize $s\mathcal{O}_{\mathbf{p}, \mathbf{q}}(\mathbf{k})$. This allows to  steer the properties of the quantum trajectories, such as the frequency of certain measurement outcomes and interestingly also their correlations. As we demonstrate below, it is possible to devise an appropriate system-ancilla interaction such that the so far artificially constructed reweighted ensemble $P(\mathbf{k},s)$ becomes the physical ensemble of a bona-fide collision-model dynamics, which can be implemented on a quantum computer.



\textit{Non-interacting ``energy"-function.---} 
We first consider a simple ``energy"-function, which solely contains external fields, $\mathbf{p} = \left( p_1, \dots, p_N \right)$, i.e., 
\begin{align}
    \mathcal{O}_\mathbf{p}(\mathbf{k}) = \sum_{n=1}^N p_n  k_n = \mathbf{p} \cdot \mathbf{k} \, . \label{eq:LocalFieldsEnergyFunction}
\end{align}
To generate the reweighted ensemble $P(\mathbf{k},s)$ we exploit an auxiliary quantum map, the so-called tilted Kraus map, which biases the probability of each quantum trajectory through the desired factor $\mathrm{e}^{-s\mathcal{O}_{\mathbf{p}}(\mathbf{k})}$. This tilted map reads, for a single time step, as
\begin{align*}
    \mathcal{E}_{s_n}[\rho] = {K}_0 \rho K_0^\dagger + \mathrm{e}^{-s_n} {K}_1 \rho K_1^\dagger \, ,
\end{align*}
with $s_n = s p_n$. While assigning the correct weights to the different trajectories, this map does not implement a physical dynamics since, for $s_n\neq 0$, it violates trace-preservation \cite{Garrahan2010,Carollo2018,Cilluffo2021}.
Nonetheless, a physical process described by the reweighted ensemble \cite{jack2010} can be constructed \cite{SM}. It consists of a map with, in general, time-dependent (i.e.~dependent on the collision number $n$) Kraus operators:
\begin{align}
    \label{eq:KrausNonInteracting0}
    \tilde{K}_0^n&= G_n K_0 G_{n-1}^{-1} \, ,\\
    \label{eq:KrausNonInteracting1}
    \tilde{K}_1^n&= \mathrm{e}^{-s_n / 2} G_n K_1 G_{n-1}^{-1}
\end{align}
for the $n$th collision. Moreover, the initial state needs to be rotated according to
\begin{align}
    R_i \ket{\psi_0} = \frac{G_0 \ket{\psi_0}}{\left\|G_0 \ket{\psi_0}\right\|}\, . \label{eq:initial_rotation}
\end{align}
The hermitian matrices ${G_n}$, are given recursively as
\begin{align}
    \label{eq:RecurrsionForNonInteracting}
    G_{n-1} = \sqrt[]{\mathcal{E}_{s_n}^*[G_n^2]}\, ,
\end{align}
using the dual tilted Kraus map $\mathcal{E}_{s_n}^*$ and setting the final-time condition $G_N = \mathds{1}$ (see Ref.~\cite{SM} for details).

\begin{figure}[t]
    \centering
    \includegraphics{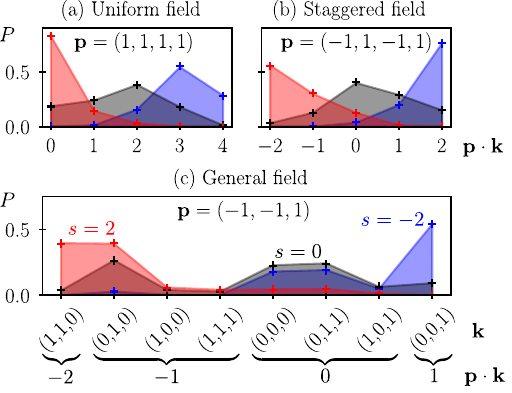}
    \caption{\textbf{Non-interacting ``energy"-function.}
    Shown is the probability $P$ for obtaining a trajectory with a given value of $\mathbf{p} \cdot \mathbf{k}$. The black data corresponds to the original, i.e. unbiased, process ($s=0$), while the red (blue) data corresponds to the biased process with $s=2$ ($s=-2$). Each data set contains two subsets: the lines above the shaded areas are numerically exact results, obtained from computing the probabilities $P(\mathbf{k}) = \left\| K_{k_N} \dots K_{k_1} \ket{\psi_0} \right\|^2$ and reweighting them appropriately. The crosses are obtained by classically simulating $20000$ trajectories generated from the biased Kraus map, Eqs.\,(\ref{eq:KrausNonInteracting0},\ref{eq:KrausNonInteracting1}). The statistical error is smaller than the marker size. In panel (c) we show the probability  as a function of $\mathbf{p} \cdot \mathbf{k}$ and also at the level of individual trajectories $\mathbf{k}$.}
    \label{fig:fig2} 
\end{figure}

In the following we discuss a few examples, for which the corresponding data are shown in Fig.~\ref{fig:fig2}.
The simplest case [panel (a)] is that of a uniform field, $p_n = 1$. According to the sign of $s$, this choice increases or decreases the probability of quantum trajectories according to the total amount of $k_n = 1$ measurement outcomes they contain [see also Fig.~\ref{fig:fig1}(b)]. 
Conversely, with a staggered field [panel (b)] $p_n = (-1)^n$ we can bias quantum trajectories towards an imbalance between odd and even spins, i.e., showing alternating measurement outcomes at odd and even times.
An even more general case [panel (c)] is that of a random sequence of local external fields, i.e., $p_n \in \{\pm  1\}$. 
In Fig.~\ref{fig:fig2} we show numerical results obtained for all three cases. The black data show the probabilities of the original, i.e., unbiased, dynamics. Throughout, we observe that for a positive value of $s$ (red data) the probability of trajectories that minimize the scalar product $\mathbf{p} \cdot \mathbf{k}$, i.e., the ``energy"-function (\ref{eq:LocalFieldsEnergyFunction}), is enhanced. Conversely, for negative $s$ trajectories that maximize $\mathbf{p} \cdot \mathbf{k}$ become dominant.

\begin{figure}[ht]
    \centering
    \includegraphics{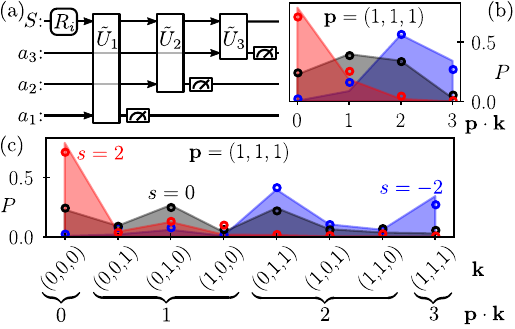}
    \caption{\textbf{Quantum simulations of the biased dynamics for a uniform field.}
    (a) The quantum circuit used to implement the biased dynamics consists of an initial rotation $R_i$ applied to the system $S$ as well as adapted collision unitaries $\tilde{U}_n$.
    Arrows indicate which qubits collide.
    The transpiled version of this quantum circuit is run on the IBM quantum processor \texttt{ibmq\_jakarta}. 
    The obtained data is evaluated with respect to (b) the ``energy" $\mathbf{p} \cdot \mathbf{k}$ and (c) the individual trajectories $\mathbf{k}$. We used 20000 samples.} 
    \label{fig:fig3}
\end{figure}

In the following we show how the biased dynamics can be implemented on an actual digital quantum processor. To do this, we need first to unravel the Kraus map defined by the operators $\{\tilde{K}^n_{k_{n}}\}^1_{k_n=0}$ into a unitary collision model with an auxiliary ancilla \cite{Leontica2021}. 
The existence of such a unitary is guaranteed by the Stinespring dilation theorem \cite{Stinespring1955,Friedman2022}.
The desired Kraus operators are obtained when choosing for the $n$th collision the operator (see Ref.~\cite{SM} for details)
\begin{align}
    \tilde{U}_n = \left({\begin{array}{c|c}
        \tilde{K}_0^n &  \dots\\
        \tilde{K}_1^n &  \dots
    \end{array}}\right) \, .
    \label{eq:StinespringDilation}
\end{align}
The rectangular submatrix containing $\tilde{K}_0^n$ and $\tilde{K}_1^n$ is an isometry, since  $\tilde{K}_{0}^{n\, \dagger} \tilde{K}_{0}^n+\tilde{K}_{1}^{n\, \dagger} \tilde{K}_{1}^n=\mathds{1}$. The other columns do not participate to the collision-model dynamics and we can thus fill them with additional orthogonal column vectors. In this way, the isometry is promoted to a unitary operator encoding the system-ancilla collision.  
This procedure allows us to construct the quantum circuit shown in Fig.~\ref{fig:fig3}(a), which must be transpiled using the gates available for the chosen quantum processor, a task which is left unsupervised to the Qiskit library \cite{Qiskit2019}. 

In Fig.~\ref{fig:fig3}(b,c), we show results from quantum simulations of our trajectories on the $7$-qubit quantum processor \texttt{ibmq\_jakarta}.  
The obtained probabilities for the original and the modified trajectory ensemble (circles) show good agreement with the exact probabilities (shaded), both as a function of the scalar product $\mathbf{p}\cdot \mathbf{k}$ and at the level of individual trajectories $\mathbf{k}$. In contrast to the classical simulation results, displayed in Fig.~\ref{fig:fig2}, some deviations can be observed, that exceed the statistical error bounds (which are smaller than the marker size). This points towards the presence of a systematic error in the quantum processor due to a noisy implementation of the unitary gates \cite{Leymann2020,Preskill2018,Lau2022,DiBartolomeo2023,Liu2020,Dahlhauser2021}.



\textit{Interacting ``energy"-function.---} We now turn to the more interesting case of interacting ``energy"-functions, which permit the bias of time correlations  among ancillary measurements within quantum trajectories. For concreteness, we consider here the following nearest-neighbor function
\begin{align}
    \mathcal{O}_\text{NN}(\mathbf{k}) = \sum_{n=2}^N \sigma_n \sigma_{n-1} \, ,
    \label{eq:NearestNeighborEnergyFunction}
\end{align}
with $\sigma_n = 1-2k_n$. In our thermodynamic analogy, this function corresponds to a classical Ising energy and large values of $\mathcal{O}_\text{NN}(\mathbf{k})$ are associated with microstates in which neighboring spins are aligned, i.e., measurements on consecutive ancillas yield the same outcome. 
Compared to non-interacting ``energy"-functions, this case is more challenging to treat since the biasing at a given discrete-time step $n$ depends on the outcome of the measurement of the previous ancilla's state. The ensuing tilted map needs to involve all collisions and can no longer be split into independent maps. It reads (see Ref.~\cite{SM} for details) 
\begin{equation*}
     \rho \to \sum_{\mathbf{k}\in \{0,1\}^N}  (T_s)_{k_N, k_{N-1}}\circ \dots \circ (T_s)_{k_2, k_{1}} \circ (T_s)_{k_{1}}\left[\rho \right] \, ,
\end{equation*}
where 
\begin{equation}
    (T_s)_{k_n, k_{n-1}} [\rho] = \mathrm{e}^{-s\sigma_n\sigma_{n-1}} {K}_{k_n} \rho {K}_{k_n}^\dagger
    \label{eq:TransferMatrixNearestNeighborEnergyFunction}
\end{equation}
and $(T_s)_{k_{1}}[\rho]={K}_{k_1} \rho {K}_{k_1}^\dagger$. The matrices $T_s$ are transfer matrices with  entries given by maps rather than numbers. 
As for the non-interacting case, the tilted map cannot describe a physical process since it is not trace-preserving for $s\neq0$. 
Nonetheless, an actual physical process that creates the reweighted ensemble can be found also in this case of interacting ``energy"-functions. For the $n$th collision this is defined by the Kraus operators $\{\tilde{K}^n_{k_n|k_{n-1}}\}^1_{k_n=0}$ \textit{conditioned} on the outcome $k_{n-1}$ of the measurement at time $n-1$ and given by
\begin{align}
    \label{eq:KrausNearestNeighborInteraction}
    \tilde{K}^{ n}_{k_n|k_{n-1}} = \mathrm{e}^{-s\sigma_n\sigma_{n-1} / 2} G_{n|k_n} K_{k_n} G_{n-1|k_{n-1}}^{-1}\, ,
\end{align}
with 
\begin{align}
\label{eq:RecursionRelationForNearestNeighborInteraction}
    G_{n-1|k_{n-1}} = \sqrt[]{\sum_{k_n=0}^1 (T_s)^*_{k_n, k_{n-1}}\left[G_{n|k_n}^2\right]}\, ,
\end{align}
setting $G_{N|k_{N}} = \mathds{1}$ for all $k_N$.
We note, that $G_{0|0}=G_{0|1}=G_{0}$ and the initial state is modified according to the rotation given by Eq.\,(\ref{eq:initial_rotation}). This construction can be extended to more general ``energy"-functions, for example functions which depend on strings of measurement outcomes, $k_1k_2k_3\dots k_N$. This allows to  modify any $n$-time correlation function of the trajectory ensemble.


\begin{figure}[ht]
    \centering
    \includegraphics{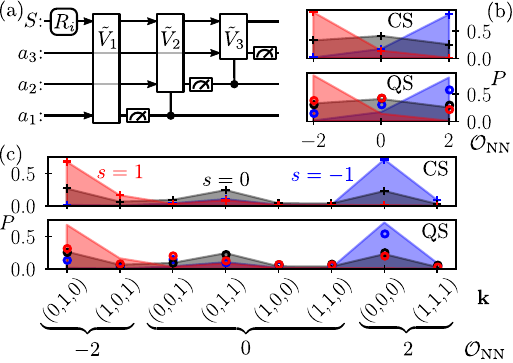}
    \caption{\textbf{ Biasing correlations within quantum trajectories.} (a) Quantum circuit for the biased dynamics with the interacting ``energy"-function $\mathcal{O}_\text{NN}$ in Eq.\,(\ref{eq:NearestNeighborEnergyFunction}). Here, $R_i$ represents the required rotation of the initial state and the $\tilde{V}_{n>1}$ are conditioned unitaries (see main text). (b,c) Probability for obtaining trajectories with given $\mathcal{O}_\text{NN}$ and $\mathbf{k}$, respectively. The black data correspond to the original, i.e. unbiased, process ($s=0$), while the red (blue) data correspond to the biased process with $s=1$ ($s=-1$). Each data set contains two subsets: the lines above the shaded areas are numerically exact results, obtained from the (tilted) dynamical map. Top panels (CS) are obtained by classically simulating $20000$ trajectories of the biased dynamics. Bottom panels (QS) are obtained by sampling the same amount of trajectories on the quantum processor \texttt{ibmq\_jakarta}.} 
    \label{fig:fig4}
\end{figure}

To implement this conditioned non-Markovian biased dynamics on a quantum processor, the first step is once again to promote each conditioned Kraus map, given by the operators in Eq.\,(\ref{eq:KrausNearestNeighborInteraction}), into a unitary collision, $\tilde{U}_{n|k_{n-1}}$, with an ancilla, in analogy to Eq.\,(\ref{eq:StinespringDilation}). Then, we introduce the projector on the state of the ancilla $\ketbra{k_{n-1}}$ at time step $n-1$ and construct the three-qubit unitary 
\begin{align*}
    \tilde{V}_n = \sum_{k_{n-1}} \ketbra{k_{n-1}} \otimes \tilde{U}_{n|k_{n-1}}\, ,
\end{align*}
which implements the collision $\tilde{U}_{n|k_{n-1}}$ on the system and on the $n$th ancilla, according to the outcome of the measurement $k_{n-1}$ on the previous ancilla. The structure of these unitaries further highlights the non-Markovian character of the biased dynamics,  which thus takes the form of an {\it extended collision model}. The corresponding quantum circuit is shown in Fig.~\ref{fig:fig4}(a).

In Figs.\,\ref{fig:fig4}(b,c) we report results from both classical numerical simulation (crosses) and the quantum simulations (circles) of this biased dynamics. As anticipated, the classical simulation results, obtained from trajectory sampling, agree excellently with the reweighted probabilities. However, the probabilities sampled via the quantum simulations display significant systematic errors, both when plotted as function of $\mathcal{O}_\text{NN}$ and when resolved for individual trajectories $\mathbf{k}$. We attribute these errors to the substantial depth of the transpiled circuit, which is mainly due to the representation of the three-qubit gates in terms of the basis gates of the device  \cite{Leontica2021}: even small single-gate errors, as for instance the \texttt{CNOT}-error which is smaller than $1\,\%$ according to the most recent calibration, accumulate and give rise to an effective dynamics which is rather different from the desired one \cite{Smith2019}.
To our understanding, simple error mitigation techniques regarding measurement errors \cite{GarciaPerez2020} or zero-noise extrapolation \cite{Keenan2022} are not sufficient to eliminate the  noise introduced in the dynamics, at least not when using the unsupervised transpiling of the circuit.  In Ref.~\cite{SM} we discuss how Qiskit allows for an a priori estimate on the errors to be expected from the quantum  simulations.


\textit{Conclusions.---} 
We have shown how to bias the dynamical behavior, e.g., the temporal correlations, of the output of open quantum systems on a quantum computer. Our approach is based on a thermodynamic analogy, where the probability of a given trajectory is  modified through  an ``energy"-function. Simple, i.e., non-interacting, ``energy"-functions can be rather reliably implemented. However, the inclusion of interactions in order to enhance correlations  between output measurements  on the ancillae  appears to exhaust the capability of the \texttt{ibmq\_jakarta} machine, on which we conducted our study. The ultimate reason appears to be that the interacting case requires the implementation of three-body gates, which are transpiled into relatively deep circuits. This shows that already the quantum simulation and manipulation of short trajectories of a two-level system is a difficult task and illustrates the enormous challenge of faithfully simulating and  engineering open many-body quantum dynamics with quantum computers. Advances in circuit design, such as dynamic circuits that allow \textit{in situ} feedback based on measurement results, may help to improve this situation \cite{Cross2022,McKay2018}. Also, the ability of resetting and reusing ancillas will of course allow to implement more collisions and thus longer evolution times. So far, we calculated the quantum circuit yielding the biased dynamics in advance on a classical computer. In the future it would be interesting, to investigate whether this is indeed necessary or whether the corresponding maps can be directly computed and implemented on a quantum device.

The code and the data that support the findings of this Letter are available on Zenodo \cite{ZenodoData}.

\acknowledgments
\textit{Acknowledgements.---} We thank Markus M\"uller, Juan P.~Garrahan and Hugo Touchette for fruitful discussions. We acknowledge funding from the Deutsche Forschungsgemeinschaft (DFG, German Research Foundation) under Project No. 435696605 and through the Research Unit FOR 5413/1, Grant No. 465199066, and through the Research Unit FOR 5522/1, Grant No. 499180199. This project has also received funding from the European Union’s Horizon Europe research and innovation program under Grant Agreement No. 101046968 (BRISQ). F.C.~is indebted to the Baden-W\"urttemberg Stiftung for the financial support of this research project by the Eliteprogramme for Postdocs.

\bibliography{biblio}

\onecolumngrid
\newpage

\setcounter{equation}{0}
\setcounter{page}{1}

\setcounter{figure}{0}
\setcounter{table}{0}
\makeatletter
\renewcommand{\theequation}{S\arabic{equation}}
\renewcommand{\thefigure}{S\arabic{figure}}
\renewcommand{\thetable}{S\arabic{table}}

\makeatletter
\renewcommand{\theequation}{S\arabic{equation}}
\renewcommand{\thefigure}{S\arabic{figure}}
\renewcommand{\thetable}{S\arabic{table}}

\begin{center}
{\Large SUPPLEMENTAL MATERIAL}
\end{center}
\begin{center}
\vspace{0.8cm}
{\Large Thermodynamics of quantum trajectories on a quantum computer}
\end{center}
\begin{center}
Marcel Cech$^{1}$, Igor Lesanovsky,$^{1,2}$ and Federico Carollo$^1$ 
\end{center}
\begin{center}
$^1${\em Institut f\"ur Theoretische Physik, Universit\"at T\"ubingen,}\\
{\em Auf der Morgenstelle 14, 72076 T\"ubingen, Germany}\\
$^2${\em School of Physics and Astronomy and Centre for the Mathematics and Theoretical Physics of Quantum Non-Equilibrium Systems, The University of Nottingham, Nottingham, NG7 2RD, United Kingdom}
\end{center}


\section{I. Continuous-time limit of the dynamics}
In this section, we show how, within an appropriate scaling limit, the collision dynamics defined in Eq.\,(\ref{eq:OriginalCollisionHamiltonian}) coincides, for the system of interest, with the continuous-time Lindblad time evolution of a driven two-level atom with decay. As briefly mentioned in the main text, the idea is to introduce a time interval $\Delta t$, which represents the duration of a single collision, and to redefine $\omega \to \omega \Delta t$ and $\kappa \to \sqrt[]{{\kappa}{\Delta t}}$. In this case, the $n$th collision is described by the unitary $U = \mathrm{e}^{-i H \Delta t}$ with
\begin{equation*}
    H = \left(\mathds{1} \otimes \omega \sigma_x\right) +\sqrt{\frac{\kappa}{\Delta t}} \left( \sigma_+ \otimes \sigma_- + \sigma_- \otimes \sigma_+ \right) = \mathds{1} \otimes H_S + H_\text{int} \, ,
\end{equation*}
where we separated the free Hamiltonian $H_S$ for the system and the interaction Hamiltonian $H_\text{int}$. In the small $\Delta t$ limit, we can expand the unitary and, evaluating the trace over the ancilla subspace (denoted as $\operatorname{Tr}'$), we find the increment for the density matrix of the system as \cite{Ciccarello2022}
\begin{equation*}
    \Delta \rho = \Delta t \left( -i \Big[H_S + \operatorname{Tr}'\left\{ H_\text{int}(\eta\otimes \mathds{1}) \right\}, \rho \Big] + \Delta t \operatorname{Tr}'\left\{ H_\text{int} (\eta\otimes \rho) H_\text{int} - \frac{1}{2} \left\{ H_\text{int}^2, \eta\otimes \rho \right\} \right\} \right) \, .
\end{equation*}
Here, we consider the state $\eta= \ketbra{0}$ for the ancillary system, which yields
\begin{align*}
    \operatorname{Tr}'\left\{ H_\text{int}(\eta\otimes \mathds{1}) \right\} &= 0 \, , \\
    \Delta t \operatorname{Tr}'\left\{ H_\text{int} (\eta\otimes \rho) H_\text{int} - \frac{1}{2} \left\{ H_\text{int}^2, \eta\otimes \rho \right\} \right\} &= \kappa \left( \sigma_- \rho \sigma_+ - \frac{1}{2} \left\{ \sigma_+ \sigma_-, \rho \right\} \right) \, .
\end{align*}
In the continuous-time limit $\Delta t \to 0$, this results in the Lindblad master equation $\dot{\rho} = \mathcal{L}[\rho]$ with
\begin{equation*}
    \mathcal{L}[\rho] = -i \omega \left[ \sigma_x, \rho \right] + \kappa \left( \sigma_- \rho \sigma_+ - \frac{1}{2} \left\{ \sigma_+ \sigma_-, \rho \right\} \right)
\end{equation*}
describing the open-system dynamics of a two-level system with Rabi frequency $\omega$ and decay rate $\kappa$ \cite{Garrahan2010}.

\section{II. Biased dynamics for non-interacting ``energy"-functions}
In this section, we briefly describe how to achieve the physical dynamics generating the reweighted ensemble of probabilities biased with respect to non-interacting ``energy"-functions. We start by writing the tilted dynamics over all collisions 
\begin{align*}
    \rho_N^s = \mathcal{E}_{s_N} \circ \dots \circ \mathcal{E}_{s_n} \circ \dots \circ \mathcal{E}_{s_1}\left[\rho_0\right]\, ,
\end{align*}
with $\rho_0  = \ketbra{\psi_0}$ \cite{Cilluffo2021} and with $\circ$ denoting composition of maps.
It is worth noticing that $\operatorname{Tr}\left\{ \rho_N^s \right\}$ gives the moment generating function of the average ``energy" $\left< \frac{\mathcal{O}_\mathbf{p}(\mathbf{k})}{N} \right>_\mathbf{k}$ \cite{Garrahan2010,Carollo2018,Cilluffo2021}. Following Ref.~\cite{Carollo2018}, we introduce the maps $g_n[X] = G_n X G_n$ and their inverse $g_n^{-1}[X]=G_n^{-1}XG_n^{-1}$, i.e., the maps for which $g_n\circ g_n^{-1}$ is the identity map. Introducing identity maps in the form of $g_n^{-1}\circ g_n$ on the left and on the right of every single time step tilted operators we can rewrite the above map as 
\begin{align*}
    \rho_N^s = g_N^{-1} \circ \left( \prod_{n = 1}^N g_n \circ \mathcal{E}_{s_n} \circ g_{n-1}^{-1} \circ \right) g_0\left[ \rho_0 \right]\, .
\end{align*}
We now exploit the gauge freedom introduced through the hermitian matrices $G_n$ in such a way that each $\Tilde{\mathcal{E}}_{n} = g_n \circ \mathcal{E}_{s_n} \circ g_{n-1}^{-1}$ in the above ordered product is trace-preserving (see, e.g., also \cite{Cilluffo2021}). Imposing the final condition $G_N = \mathds{1}$, we find the recursion relation in Eq.\,(\ref{eq:RecurrsionForNonInteracting}). The matrices $G_n$ can then be used to construct the Kraus operators in Eqs.\,(\ref{eq:KrausNonInteracting0},\ref{eq:KrausNonInteracting1}) of the map $\Tilde{\mathcal{E}}_{n}$. The initial rotation emerges from normalizing $g_0\left[ \rho_0 \right]$. In this way, the not trace-preserving character of the tilted dynamics has been transformed into an irrelevant normalization of the initial state while the tilted map has been transformed into a bona-fide quantum dynamics. 


\section{III. Recovering the biased Kraus map from Eq.\,(\ref{eq:StinespringDilation})}
In this section, we demonstrate the equivalence between the dynamics described by the unitary $\tilde{U}_n$ in Eq.\,(\ref{eq:StinespringDilation}) and the Kraus operators $\{\tilde{K}^n_{k_{n}}\}^1_{k_n=0}$ defined in Eqs.\,(\ref{eq:KrausNonInteracting0},\ref{eq:KrausNonInteracting1}) for the system of interest. We start by noticing that the joined state of system plus ancilla before the collision is given by $\ketbra{0} \otimes \rho$. Therefore, the state just after the collision, but before the measurement, is given by
\begin{align*}
    \tilde{U}_n \left( \ketbra{0}{0} \otimes \rho \right) \tilde{U}_n^\dagger = \ketbra{0} \otimes \tilde{K}_{0}^n\rho\tilde{K}_{0}^{n\, \dagger} + \ketbra{1}{0} \otimes \tilde{K}_{1}^n\rho\tilde{K}_{0}^{n\, \dagger} + \ketbra{0}{1} \otimes \tilde{K}_{0}^n\rho\tilde{K}_{1}^{n\, \dagger} + \ketbra{1} \otimes \tilde{K}_{1}^n\rho\tilde{K}_{1}^{n\, \dagger} \, ,
\end{align*}
where we inserted Eq.\,(\ref{eq:StinespringDilation}) as $\tilde{U}_n = \ketbra{0} \otimes \tilde{K}_{0}^n + \ketbra{1}{0} \otimes \tilde{K}_{1}^n + \dots$ and used the orthogonality of $\ket{0}$ and $\ket{1}$. Tracing over the ancilla, i.e., averaging over both possible measurement outcomes, results in the Kraus map $\tilde{\mathcal{E}}_n[\rho] = \tilde{K}_{0}^n\rho\tilde{K}_{0}^{n\, \dagger} + \tilde{K}_{1}^n\rho\tilde{K}_{1}^{n\, \dagger}$ described in Eqs.\,(\ref{eq:KrausNonInteracting0}, \ref{eq:KrausNonInteracting1}).

\section{IV. Transfer matrix formalism and biased dynamics for interacting ``energy"-functions}
We provide details on the derivation of the biased dynamics for the case of the nearest-neighbour ``energy"-function in Eq.\,(\ref{eq:NearestNeighborEnergyFunction}). We start by writing the tilted map
\begin{equation}
    \begin{split}
        \rho_N^s = 
        & \sum_{\mathbf{k}\in \{0,1\}^N} \mathrm{e}^{-s\sigma_N \sigma_{N-1}} K_{k_N} \dots \mathrm{e}^{-s\sigma_2 \sigma_{1}} K_{k_2} K_{k_1} \rho K_{k_1}^\dagger  K_{k_2}^\dagger \dots K_{k_N}^\dagger \\
        =& \sum_{\mathbf{k}\in \{0,1\}^N}  (T_s)_{k_N, k_{N-1}}\circ \dots \circ (T_s)_{k_2, k_{1}} \circ (T_s)_{k_{1}}\left[\rho_0 \right] \, ,
        \label{eq:DeformedDynamicsNearestNeighborEnergyFunction}
    \end{split}
\end{equation}
where we expanded  $\mathrm{e}^{-s\mathcal{O}_\text{NN}(\mathbf{k})}$ into the product of the different exponentials $\mathrm{e}^{-s\sigma_n \sigma_{n-1}}$ and grouped these reweighting factors together with the Kraus operators to introduce the transfer matrices from Eq.\,(\ref{eq:TransferMatrixNearestNeighborEnergyFunction}). Following the approach for non-interacting ``energy"-functions, we introduce the conditioned maps $g_{n|k_n}[\rho] = G_{n|k_n} \rho G_{n|k_n}$ and their inverse $g_{n|k_n}^{-1}$. We highlight that we can write $g_n$ defined by $(g_n)_{k_n, k_n'}[\rho] = \delta_{k_n, k_n'} g_{n|k_n} [\rho]$ as a block-diagonal matrix with entries given by maps in close  resemblance with  the form of the transfer matrices $T_s$. Thus, Eq.\,(\ref{eq:DeformedDynamicsNearestNeighborEnergyFunction}) yields
\begin{align}
    \rho_N^s = \sum_{\mathbf{k}\in \{0,1\}^{N}} g_{N|k_N}^{-1} \circ \left( \prod_{n = 1}^N g_{n|k_n} \circ (T_s)_{k_n, k_{n-1}} \circ g_{n-1|k_{n-1}}^{-1} \circ \right) g_{0}[\rho_0]\, . 
    \label{eq:GaugeTransformedDeformedDynamicsNearestNeighborEnergyFunction}
\end{align}
We identify the \textit{conditioned} maps $\Tilde{\mathcal{E}}_{n|k_{n-1}} = \sum_{k_n} g_{n|k_{n}} \circ (T_s)_{k_n, k_{n-1}} \circ g_{n-1|k_{n-1}}^{-1}$. We fix the conditioning $k_{n-1}$ and impose trace-preservation by $\Tilde{\mathcal{E}}_{n|k_{n-1}}^*[\mathds{1}] = \sum_{k_n} g_{n-1|k_{n-1}}^{-1} \circ (T_s)_{k_n, k_{n-1}}^* \circ g_{n|k_n} [\mathds{1}] = \mathds{1}$ using the dual maps marked by the superscript $^*$ \cite{Cilluffo2021}. Applying $g_{n-1|k_{n-1}}$ to both sides of the equation results in the recursion relation for the hermitian, conditioned matrices $G_{n|k_n}$ [see Eq.\,(\ref{eq:RecursionRelationForNearestNeighborInteraction})]. The conditioned Kraus operators in Eq.\,(\ref{eq:KrausNearestNeighborInteraction}) are now extracted from $\Tilde{\mathcal{E}}_{n|k_{n-1}}$.

We note, that we implicitly assumed $g_{0|0} = g_{0|1} = g_{0}$ in Eq.\,(\ref{eq:GaugeTransformedDeformedDynamicsNearestNeighborEnergyFunction}). To justify this assumption, we notice the last step of the recursion relation, starting with $G_{N|k_N} = \mathds{1}$ for all $k_N$, $G_{0|k_0} = \sqrt{\sum_{k_1} (T_s)_{k_1}^*[G_{1|k_1}^2]}$ is independent of a possible $k_0$ resulting directly in $G_{0|0}=G_{0|1}=G_{0}$. This uniquely defines the initial rotation applied to the system as well as the first collision described by $\Tilde{V}_1 = \Tilde{U}_{1|0} = \Tilde{U}_{1|1}$.

\section{V. A Priori analysis of Quantum-simulation Errors}

In this section, we briefly discuss how the Qiskit library can  be used to perform an a priori estimate of the error that can be expected when running simulations on the actual quantum computer. 

\begin{figure}
    \centering
    \includegraphics{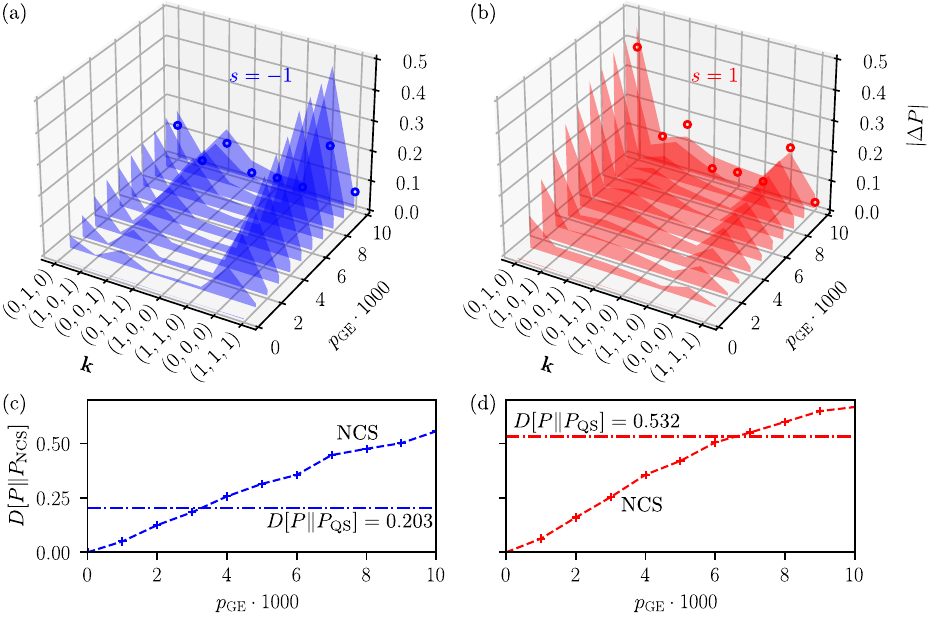}
    \caption{\textbf{Error analysis via classical noisy simulations.} (a,b) We show the absolute difference $|\Delta P|$ between the  numerically exact probability $P({\bf k},s)$ and the same obtained via classical noisy simulations (lines above shaded areas) of our biasing approach, $P_{\rm NCS}$, or the quantum simulations from Fig.~\ref{fig:fig4} (circles), $P_{\rm QS}$. The classical noisy simulations consist of a simplified noise model with readout error $p_\mathrm{R} = 0.02$ and gate error $p_\mathrm{GE}$. (c,d) We further calculate the Kullback-Leibler divergence $D[P\|P_{\rm NCS}]$ in Eq.\,(\ref{eq:KL-div}) between the exact probability and the one obtained via the classical noisy simulations (crosses). The horizontal lines provide the reference Kullback-Leibler divergence $D[P\|P_{\rm QS}]$ between exact probabilities and those obtained via quantum simulation of biased dynamics, $P_{\rm QS}$, in Fig.~\ref{fig:fig4}.}
    \label{fig:fig_s1}
\end{figure}

\noindent The main idea is that, exploiting Qiskit, one can straightforwardly define a noise model in which the type of noises and their strengths can be controlled. The noise model can then be used to simulate how different errors influence the outcome of the quantum computation. Here, for concreteness, we consider a simple noise model where we allow for a readout error $p_\mathrm{R}$ involving the measurement outcome as well as for a gate error $p_{\rm GE}$ affecting each gate in the transpiled quantum circuit. We model both errors by utilizing the bit-flip channel with the respective probability \cite{Nielsen2010}. In this way, each ancilla measurement and applied gate is accompanied by the application of this noise channel on the respective qubits \cite{QiskitNoiseModel2023}. With the noise model, we can sample the probability $P_{\rm NCS}({\bf k},s)$ through noisy classical simulations and compare it with the numerically exact probability distribution $P({\bf k},s)$. \\ 

\noindent In Fig.~\ref{fig:fig_s1}(a,b), we show the difference $|\Delta P|=|P_{\rm NCS}({\bf k},s)-P({\bf k},s)|$ for two different values of $s$. 
As expected, we see that the error increases when  considering larger gate errors, $p_{\rm GE}$. It further appears that small readout errors $p_{\rm R}$ have a less dramatic impact. This is expected since the readout is only conducted once per ancilla. As a further quantitative measure for the difference between $P_{\rm NCS}({\bf k},s)$ and $P({\bf k},s)$ we also calculate the Kullback-Leibler divergence which, given two probability distributions $Q_1({\bf k})$ and $Q_2({\bf k})$, is defined as 
\begin{equation}
    D[Q_1 \| Q_2]=\sum_{\bf k}Q_1({\bf k})\log \frac{Q_1({\bf k})}{Q_2({\bf k})}\, .
    \label{eq:KL-div}
\end{equation}
This quantity displays a similar behavior as $|\Delta P|$, as shown in Fig.~\ref{fig:fig_s1}(c,d).

\begin{figure}
    \centering
    \includegraphics{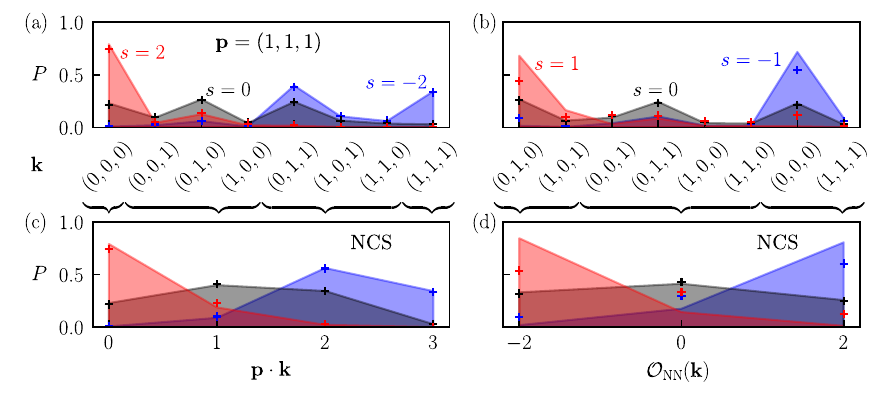}
    \caption{\textbf{Error analysis using the calibrated noise model of \texttt{ibmq\_jakarta}.} We compare the exact numerical probabilities (lines above shaded areas) to the empirical probability distribution (crosses) obtained from classical simulations, $P_{\rm NCS}$, utilizing the noise model of the latest calibration of the \texttt{ibmq\_jakarta} quantum device. The data is evaluated with respect to the ``energy'' (top panels) and the individual trajectories (bottom panels). Panels (a,c) represent the results with respect to the uniform field (see Fig.~\ref{fig:fig3}), whereas panels (b,d) represent the results of the nearest-neighbor ``energy''-function (see Fig.~\ref{fig:fig4}).}
    \label{fig:fig_s2}
\end{figure}

More importantly, Qiskit also allows to run simulations with a noise model, which exploits the latest calibration data of a considered quantum machine. This allows one to assess in advance the performance of the quantum simulation and allows to understand the impact of errors. In Fig.~\ref{fig:fig_s2}, we report this prediction for the biased dynamics shown in the main text. Overall, we see that the calibrated noise model rather accurately reflects the (open-system) dynamics of the quantum processor. It appears to slightly overestimate the actual performance as quantified by the Kullback-Leibler divergence [see Table \ref{tab:tab_s1}].



\begin{table}[htbp]
  \centering
  \caption{\textbf{Comparison between quantum simulations and noisy classical ones.} The table shows a comparison between the exact probability $P({\bf k},s)$, the ones sampled through the quantum processor $P_{\rm QS}({\bf k},s)$ and those obtained with the classical simulations using the calibrated noise model  $P_{\rm NCS}({\bf k},s)$. (a) Biased dynamics with respect to the uniform field [see also Fig.~\ref{fig:fig3} and Fig.~\ref{fig:fig_s2}(a,c)]. The different columns provide the values of Kullback-Leibler divergence $D[P\|P_{\rm QS}]$, $D[P\|P_{\rm NCS}]$ and $D[P_{\rm QS}\|P_{\rm NCS}]$, respectively. (b) Same as in (a) for the case of biased dynamics with respect to the interacting ``energy"-function [see also Fig.~\ref{fig:fig4} and Fig.~\ref{fig:fig_s2}(b,d)]. }
    \begin{tabular}{l|r|r|rrl|r|r|r}
    \multicolumn{4}{c}{(a) Uniform field} &       & \multicolumn{4}{c}{(b) Nearest-neighbor ``energy''-function}   \\
          & $D[P\|P_{\rm QS}]$ & $D[P\|P_{\rm NCS}]$ & $D[P_{\rm QS}\|P_{\rm NCS}]$ & {\hspace*{1cm}} &       & $D[P\|P_{\rm QS}]$ & $D[P||P_{\rm NCS}]$ & $D[P_{\rm QS}\|P_{\rm NCS}]$  \bigstrut[b]\\
\cline{1-4}\cline{6-9}    $s = -2$ & 0.043 & 0.002 & 0.040 &       & $s = -1$ & 0.203 & 0.128 & 0.019 \bigstrut[t]\\
    $s = 0$ & 0.019 & 0.001 & 0.016 &       & $s = 0$ & 0.020 & 0.004 & 0.013 \\
    $s = 2$ & 0.065 & 0.009 & 0.055 &       & $s = 1$ & 0.532 & 0.253 & 0.069 \\
    \end{tabular}%
  \label{tab:tab_s1}%
\end{table}%


\end{document}